\documentclass{article}
\usepackage{spconfa4,amsmath,graphicx}
\usepackage{amsfonts}
\usepackage{subcaption}
\usepackage{tikz}
\usetikzlibrary{calc,chains,shapes,positioning}
\usepackage{comment}
\usepackage{pgfplots}
\usepackage{phaistos}
\usepackage{float}
\usepackage{phaistos}
\usepackage{pgfplots}
\usepackage{cite}
\usepackage{mathdots}
\usepackage{scalerel}
\usepackage{enumitem}
\usepackage{accents}
\usepackage{balance}
\usepackage[ruled]{algorithm2e}
\SetAlgoHangIndent{0pt}
\usepackage[moderate,title=normal,sections=normal,margins=normal,bibliography=normal,mathdisplays=normal,mathspacing=normal, lists=normal]{savetrees}
\usepackage{balance}
\usepackage{epstopdf}

\newcommand{\ubar}[1]{\underaccent{\bar}{#1}}
\DeclareMathOperator*{\argmin}{argmin}

\title{Reference Microphone Selection for the weighted prediction error\\ algorithm using the normalized L-p norm}
\name{Anselm Lohmann$^{1}$, Toon van Waterschoot$^{2}$, Joerg Bitzer$^{3}$, Simon Doclo$^{1,3}$ \thanks{This work has received funding from the European Union’s Horizon 2020 research and innovation programme under the Marie Skłodowska-Curie grant
agreement No. 956369 and KU Leuven Internal Funds C14/21/075, and from the Deutsche Forschungsgemeinschaft (DFG, German Research Foundation) under Germany's Excellence Strategy - EXC 2177/1 - Project ID 390895286.}}
\address{$^{1}$Carl von Ossietzky Universität Oldenburg, Dept. of Medical Physics and Acoustics, Germany \\
$^{2}$KU Leuven, Department of Electrical Engineering (ESAT-STADIUS), Leuven, Belgium\\
$^{3}$Fraunhofer IDMT, Project Group Hearing, Speech and Audio Technology, Oldenburg, Germany \\
{\tt anselm.lohmann@uni-oldenburg.de}
}
\begin{document}
%
\maketitle
\begin{abstract}
Reverberation may severely degrade the quality of speech signals recorded using microphones in a room. For compact microphone arrays,
the choice of the reference microphone for multi-microphone dereverberation typically does not have a
large influence on the dereverberation performance. In contrast, when the microphones are spatially distributed, the choice of the reference microphone may significantly contribute to the dereverberation performance. In this paper, we propose to perform reference microphone selection for the weighted prediction error (WPE) dereverberation algorithm based on the normalized $\ell_p$-norm of the dereverberated output signal. Experimental results for different source positions in a reverberant laboratory show that the proposed method yields a better dereverberation performance than reference microphone selection based on the early-to-late reverberation ratio or signal power.
\end{abstract}
\section{Introduction}
\label{sec:intro}

Microphone recordings of a speech source inside a room are typically degraded by reverberation, i.e. acoustic reflections against walls and objects in the room. While early reflections may improve speech intelligibility, late reverberation typically reduces both speech intelligibility as well as automatic speech recognition performance \cite{Beutelmann2006,YoshiokaASR2012}. Therefore, effective speech dereverberation is required for many applications, including voice-controlled systems, hearing aids and hands-free telephony \cite{HabetsNaylor2018,Cauchi2015,Lemercier2023,Williamson2017,Nakatani2010,Jukic2015}. A popular blind multi-channel dereverberation algorithm is the weighted prediction error (WPE) algorithm \cite{Nakatani2010,Jukic2015}, which is based on multi-channel linear prediction (MCLP). WPE performs dereverberation in a chosen reference microphone by estimating the late reverberant component using a prediction filter and subtracting this estimate from the reference microphone signal. \par
When performing multi-microphone speech enhancement using compact microphone arrays, the choice of the reference microphone typically does not have a large influence on the quality of the output signal. However, when considering spatially distributed microphones, there may be large differences in the early-to-late reverberation ratio (ELR) and signal power in each microphone. Hence, the choice of the reference microphone may significantly contribute to the speech enhancement performance \cite{LawinOre2012,Zhang2021,Araki2018}. In \cite{LawinOre2012} and \cite{Zhang2021}, the reference microphone selection problem was formulated for different multi-microphone noise reduction algorithms by maximizing the output signal-to-noise ratio. In \cite{Araki2018}, different reference microphone selection methods were proposed for speech enhancement in meeting recognition scenarios when considering different microphone sensitivities. However, to the best of the authors' knowledge, no work exists on reference microphone selection for multi-microphone dereverberation. 

In this paper, we propose to perform reference microphone selection for the WPE algorithm based on the normalized $\ell_p$-norm of the dereverberated output signal. From the WPE optimization problem, it may appear logical to formulate the reference microphone selection problem as an $\ell_p$-norm minimization problem. However, since the $\ell_p$-norm depends on the signal power, which may greatly vary for spatially distributed microphones, we propose to normalize for the output signal power, leading to a selection based on the ratio of the $\ell_p$-norm and the $\ell_2$-norm \cite{Hurley2009,Li2014,Jia2017}. Experimental results for several source positions and spatially distributed microphones in a reverberant laboratory show that the dereverberation performance using the proposed reference microphone selection method is larger than the performance when selecting the reference microphone based on the estimated ELR \cite{Xiong2019} or signal power \cite{LawinOre2012}\cite{Araki2018}. Furthermore, similar performance can be achieved for the proposed method using only a small number of WPE iterations.

\section{Signal Model}
\label{sec:format}

We consider a scenario where a single speech source is captured in a  room by $M$ spatially distributed microphones. Similarly as in \cite{Nakatani2010,Jukic2015}, we consider a static scenario without additive noise. In the short-time Fourier transform (STFT) domain, let $s (f,n)$ denote the clean speech signal with $f \in \{1,...,F\}$ the frequency bin index and $n \in \{1,...,N\}$ the time frame index, where $F$ and $N$ denote the number of frequency bins and time frames, respectively. The reverberant signal at the $m$-th microphone $x_m (f,n)$ can be written as
\begin{equation} 
    x_m (f,n) = \sum_{l = 0}^{L_h - 1}h_m(f,l)s(f,n-l) + e_m (f,n),
    \label{eq:sigmodel_stft}
\end{equation}
where $h_m (f,l)$ denotes the subband convolutive transfer function with length $L_h$ between the speech source and the $m$-th microphone, and $e_m(f,n)$ denotes the subband modelling error \cite{Avargel2007}. In the remainder of the paper, the frequency bin index $f$ will be omitted where possible. Assuming the subband modelling error to be 0, the dereverberation problem in microphone $r$ (referred to as the reference microphone) can be formulated as (see Fig. 1)
\begin{equation}
    d_r(n) = x_r(n) - u_r(n).
    \label{eq:sig_model}
\end{equation}
The desired component $d_r(n) = \sum_{l = 0}^{L_d - 1}h_r(l)s(n-l)$ consists of the direct path and early reflections in the reference microphone signal $x_r(n)$, where $L_d$ denotes the temporal cut-off between early and late reflections. The undesired component $u_r(n) = \sum_{l = L_d}^{L_h - 1}h_r(l)s(n-l)$, which we aim to estimate, is the late reverberant component in the reference microphone signal $x_r(n)$. It should be noted that for spatially distributed microphones, the power of the desired component $d_r(n)$ and the power of the undesired component $u_r(n)$ may greatly depend on the choice of reference microphone $r$. \par Using the MCLP model \cite{Nakatani2010}, the late reverberant component $u_r(n)$ can be written as the sum of filtered delayed versions of all reverberant microphone signals. Whereas for compact microphone arrays the same prediction delay is typically used for each microphone, it has been shown in \cite{Lohmann2023} that for spatially distributed microphones it is beneficial to use a microphone-dependent prediction delay, i.e.
\begin{equation}
u_r(n) = \sum_{m = 1}^{M}\sum_{l=0}^{L_g - 1}g_{m,r}(l)x_m(n - \tau_{m,r} - l),
\label{eq:new_MCLP}
\end{equation}
where $g_{m,r}(l)$ denotes the $m$-th prediction filter of length $L_g$ and $\tau_{m,r}$ denotes the prediction delay for the $m$-th microphone.
Using \eqref{eq:new_MCLP}, the signal model in \eqref{eq:sig_model} can be rewritten in vector notation  as
\begin{equation}
    \mathbf{d}_r = \mathbf{x}_r - \mathbf{X}_{\boldsymbol{\tau},r}\mathbf{g}_r,
    \label{eq:new_vec_MCLP}
\end{equation}
with
\begin{equation}
\mathbf{d}_r = \begin{bmatrix} 
d_r(1)&\cdots & d_r (N)
\end{bmatrix}^{T} \in\mathbb{C}^{N},
\label{eq:d_def}
\end{equation}
\begin{equation}
\mathbf{x}_r = \begin{bmatrix} 
x_r(1)&\cdots & x_r(N) 
\end{bmatrix}^{T} \in\mathbb{C}^{N}.
\end{equation}
The multi-channel delayed convolution matrix $\mathbf{X}_{\boldsymbol{\tau},r}$ in \eqref{eq:new_vec_MCLP} is defined as
\begin{equation}
\mathbf{X}_{\boldsymbol{\tau},r} = \begin{bmatrix} 
\mathbf{X}_{\tau_{1,r}}&\cdots & \mathbf{X}_{\tau_{M,r}}
\end{bmatrix}\in\mathbb{C}^{N\times ML_g},
\end{equation} 
where $\mathbf{X}_{\tau_{m,r}} \in\mathbb{C}^{N\times L_g}$ is the convolution matrix of $\mathbf{x}_m$ delayed by $\tau_{m,r}$ frames with $\tau$ the prediction delay in the reference microphone and $\mathbf{g}_r \in \mathbb{C}^{ML_g}$ is the stacked vector of all prediction filter coefficients $g_{m,r}(l)$. The dereverberation problem, i.e. estimation of the desired component $\mathbf{d}_r$, is now reduced to estimating the filter $\mathbf{g}_r$ predicting the undesired late reverberant component.
\section{WPE Algorithm}
\begin{figure}[t!]
\hspace{1.28cm}\includegraphics[width=0.8\columnwidth]{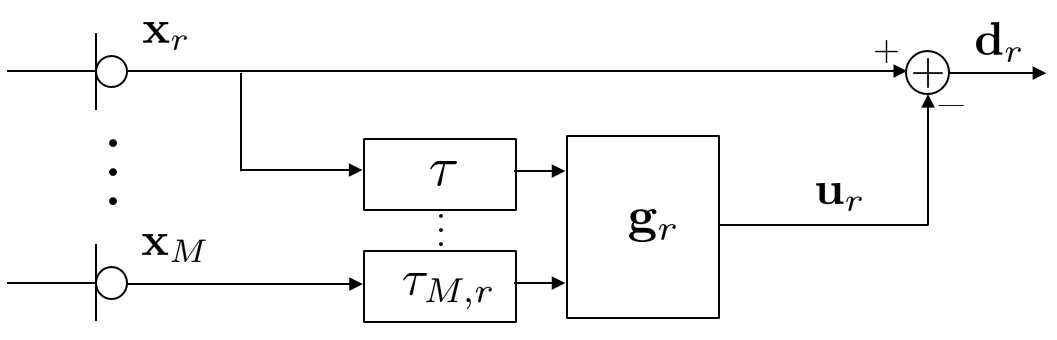}
\caption{WPE with microphone-dependent prediction delays \cite{Lohmann2023}}
\label{fig:block_diag}
\vspace{-0.6cm}
\end{figure}%
\label{sec:pagestyle}
Since the desired speech component $\mathbf{d}_r$ can be assumed to be sparser than the reverberant microphone signal $\mathbf{x}_r$, it has been proposed in \cite{Jukic2015} to compute the prediction filter $\mathbf{g}_r$ by minimizing the sparsity-promoting $\ell_p$-norm of the output in \eqref{eq:new_vec_MCLP}, i.e.
\begin{equation}
\min_{\mathbf{g}_r}J(\mathbf{g}_r) = \left\| \mathbf{d}_r\right\|_p^p = \left\| \mathbf{x}_r - \mathbf{X}_{\boldsymbol{\tau},r}\mathbf{g}_r\right\|_p^p,
\label{ref:og_problem}
\end{equation}
where the $\ell_p$-norm is defined as $\left\| \mathbf{d}_r\right\|_p = (\sum_{n=1}^{N}\lvert d_r(n)\rvert^{p})^{1/p}$. For effective dereverberation, the sparsity-promoting parameter $p$ is typically chosen in the range $0<p<1$ \cite{Jukic2015}, leading to a non-convex optimization problem in \eqref{ref:og_problem}. \par
A popular method for solving non-convex optimization problems such as \eqref{ref:og_problem} is the iteratively reweighted least-squares (IRLS) algorithm \cite{Daubechies2010}, where the original problem is replaced with a series of convex quadratic problems. Namely, in the $i$-th iteration the $\ell_p$-norm in \eqref{ref:og_problem} is approximated by a weighted $\ell_2$-norm, i.e. 
\begin{equation}
    \left\| \mathbf{d}_r\right\|_p^p \approx \left\| \mathbf{d}_r\right\|_{\mathbf{W}_r^{(i)}}^2 = \mathbf{d}_r^H\mathbf{W}_r^{(i)}\mathbf{d}_r,
    \label{ref:wl2_approx}
\end{equation}
where $\mathbf{W}_r^{(i)} = \text{diag}\left (\mathbf{w}_r^{(i)}\right)$ is a diagonal matrix of the weight vector in the $i$-th iteration $\mathbf{w}_r^{(i)}$.  
Given a previous estimate $\hat{\mathbf{w}}_r^{(i-1)}$ of the weights $\mathbf{w}_r^{(i)}$, the minimization problem in the $i$-th iteration can be written as 
\begin{equation}
    \min_{\mathbf{g}_r}\left\| \mathbf{x}_r - \mathbf{X}_{\boldsymbol{\tau},r}\mathbf{g}_r\right\|_{\hat{\mathbf{W}}_r^{(i-1)}}^2,
\end{equation}
yielding a closed-form solution for the prediction filter
\begin{equation}
    \hat{\mathbf{g}}_r^{(i)} = \left(\mathbf{X}_{\boldsymbol{\tau},r}^{H}(\hat{\mathbf{W}}_r^{(i-1)})^{-1}\mathbf{X}_{\boldsymbol{\tau},r}\right)^{-1}\mathbf{X}_{\boldsymbol{\tau},r}^{H}(\hat{\mathbf{W}}_r^{(i-1)})^{-1}\mathbf{x}_r.
    \label{ref:g_solution}
\end{equation}
The estimated weights $\hat{\mathbf{w}}_r^{(i-1)}$ are subsequently updated such that the approximation in (8) is a first-order approximation \cite{Jukic2015}, i.e.
\begin{equation}
    \hat{\mathbf{w}}_r^{(i)} = \lvert\hat{\mathbf{d}}_r^{(i)}\rvert^{2-p},
    \label{eq:sparsity_p}
\end{equation}
where the dereverberated output in the $i$-th iteration $\hat{\mathbf{d}}_r^{(i)} = \mathbf{x}_r - \mathbf{X}_{\boldsymbol{\tau},r}\hat{\mathbf{g}}_r^{(i)}$ with the $\lvert . \rvert$ and $(.)^{2-p}$ operators applied element-wise. To prevent division by zero, a small positive constant $\epsilon$ is typically included in the weight update in \eqref{eq:sparsity_p}. The initial weights $\hat{\mathbf{w}}_r^{(0)}$ are computed by defining an initial prediction filter $\hat{\mathbf{g}}_r^{(0)} = \mathbf{0}$. In total, $I_{\text{WPE}}$ iterations are performed.
\section{Reference Microphone Selection}
\label{sec:typestyle}
When considering spatially distributed microphones, the power of the desired component $\mathbf{d}_r$ and the power of the undesired component $\mathbf{u}_r$ may greatly vary depending on the reference microphone $r$. Hence, the quality of the dereverberated output $\hat{\mathbf{d}}_r$ may also depend significantly on the choice of reference microphone. Based on the WPE cost function, in Section 4.1  we first define the reference microphone selection problem as an $\ell_p$-norm minimization problem. However, since the differences in signal power between the microphones may be large, the $\ell_p$-norm-based reference microphone selection problem may not yield the dereverberated output $\hat{\mathbf{d}}_r$ with the highest signal quality. Hence, in Section 4.2 we propose to normalize for the power in the dereverberated output, leading to reference microphone selection based on the normalized $\ell_p$-norm.
\subsection{Reference microphone selection using $\ell_p$-norm}
By considering the WPE cost function in \eqref{ref:og_problem}, it may appear logical to select the reference microphone as the one minimizing the cost function  
\begin{equation}
    \min_{r} \left\|\hat{\mathbf{d}}^{(I)}_r\right\|_{p}=\left\| \mathbf{x}_r - \mathbf{X}_{\boldsymbol{\tau},r}\hat{\mathbf{g}}^{(I)}_{r}\right\|_{p},
    \label{ref:lp_reference_selection}
\end{equation}
where $\hat{\mathbf{d}}^{(I)}_{r}$ and $\hat{\mathbf{g}}^{(I)}_{r}$ correspond to the dereverberated output and the prediction filter for reference microphone $r$ after $I$ WPE iterations. Since WPE is run independently per frequency, a different reference microphone may be selected for each frequency when using \eqref{ref:lp_reference_selection}. In order to select a single reference microphone over all frequencies, we propose to minimize the sum over all frequencies, i.e. 
\begin{equation}
    \boxed{\hat{r}_{\ell_p}^{(I)} = \argmin_{r} \sum_{f=1}^{F}\left\| \mathbf{x}_r(f) - \mathbf{X}_{\boldsymbol{\tau},r}(f)\hat{\mathbf{g}}^{(I)}_{r}(f)\right\|_{p}},
    \label{ref:lp_reference_selection_broadband}
\end{equation}
where $\hat{r}_{\ell_p}^{(I)}$ denotes the selected reference microphone based on the $\ell_p$-norm. \par

\subsection{Reference microphone selection using normalized $\ell_p$-norm}
When the differences in signal power are large between the microphones, selecting the reference microphone based on the $\ell_p$-norm of the output may not yield the best dereverberated output, but possibly the output with the smallest power (irrespective of the amount of reverberation reduction). In order to normalize for the signal power in the different microphones, we propose to normalize the dereverberated output using the $\ell_2$-norm, i.e.
\begin{equation}
    \ubar{\hat{\mathbf{d}}}^{(I)}_r = \frac{\hat{\mathbf{d}}^{(I)}_r}{\lVert\hat{\mathbf{d}}^{(I)}_r\rVert_2}.
    \label{ref:power_normalization}
\end{equation}
When inserting the normalized dereverberated output $\ubar{\hat{\mathbf{d}}}^{(I)}_r$ into the problem in \eqref{ref:lp_reference_selection_broadband}, the modified reference microphone selection problem can be reformulated as a normalized $\ell_p$-norm minimization problem, i.e.
\begin{equation}
    \boxed{\hat{r}^{(I)}_{\ell_p/\ell_2} = \argmin_{r} \sum_{f=1}^{F}\frac{\left\| \mathbf{x}_r(f) - \mathbf{X}_{\boldsymbol{\tau},r}(f)\hat{\mathbf{g}}^{(I)}_{r}(f)\right\|_{p}}{\left\|\mathbf{x}_r(f) - \mathbf{X}_{\boldsymbol{\tau},r}(f)\hat{\mathbf{g}}^{(I)}_{r}(f)\right\|_{2}}},
    \label{ref:normalized_lp_norm_problem}
\end{equation}\newline
where $\lVert.\rVert_p/\lVert.\rVert_2$ and $\hat{r}^{(I)}_{\ell_p/\ell_2}$ denote the normalized $\ell_p$-norm \cite{Hurley2009,Li2014,Jia2017} and the selected reference microphone based on the normalized $\ell_p$-norm, respectively. The normalized $\ell_p$-norm, also known as the $\ell_p/\ell_q$-norm, is a popular alternative to the $\ell_p$-norm due to its scale-invariance \cite{Jia2017}. Typically, $q$ is chosen such that $q > 1$, with $q = 2$ being a common choice due to its relation to signal power \cite{Hurley2009}. 
\section{Experimental evaluation}
In this section, we evaluate the performance of the proposed WPE reference microphone selection method for spatially distributed microphones in a reverberant room. In Section 5.1, we discuss the considered acoustic scenario and algorithm parameters. In Section 5.2, we present the simulation results and evaluate the performance of the proposed method against a selection based on the ELR and signal power.
\subsection{Acoustic setup and algorithm parameters}
\begin{figure}[t!]
\centering
\includegraphics[width=0.35\columnwidth]{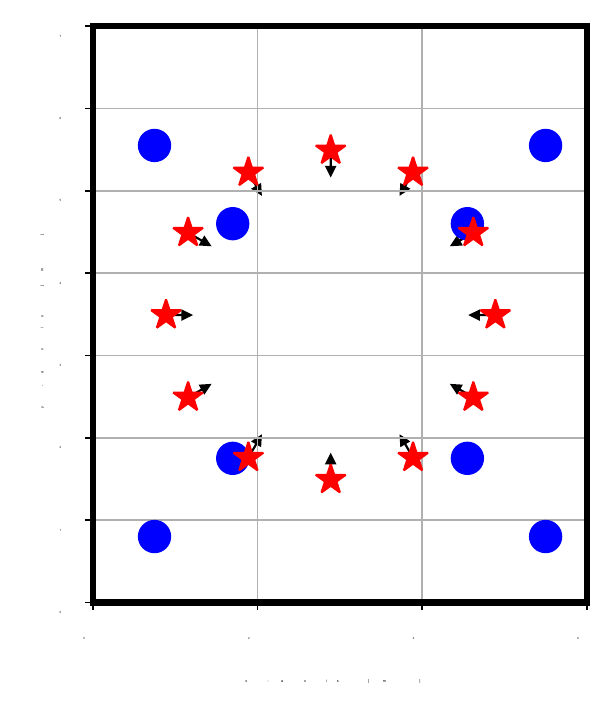}
\vspace{-0.2cm}
\caption{Positions of $M=8$ spatially distributed microphones (\includegraphics[width=0.2cm]{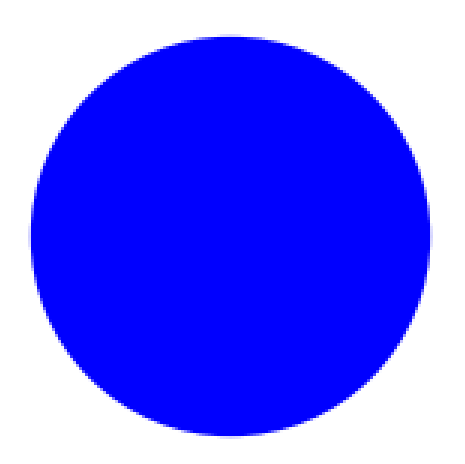}) and 12 considered speech source positions (\includegraphics[width=0.2cm]{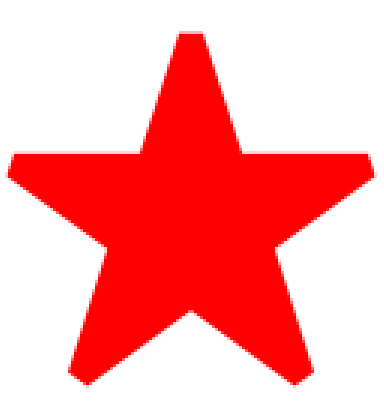})}
\label{fig:acoustic_scen}
\vspace{-0.55cm} 
\end{figure}

We consider $M = 8$ spatially distributed microphones and a single static (directional) speech source in a laboratory with dimensions of about $6$m$\times7$m$\times2.7$m and reverberation time $T_{60} \approx 1300$ ms. Fig. \ref{fig:acoustic_scen} depicts the position of the microphones and the 12 considered positions of the speech source. \par
The reverberant microphone signals were generated at a sampling rate of $16$ kHz by convolving anechoic speech signals from the TIMIT database \cite{TIMIT} with measured room impulse responses from the BRUDEX database \cite{Fejgin2023}. The signals were processed using an STFT framework with frame size of $1024$ samples, a frame shift $L_{\text{shift}} = 256$ samples and square-root Hann analysis and synthesis windows.  \par 
The WPE algorithm was run using the entire speech utterance (batch processing) and implemented with prediction filter length $L_g = 15$, number of reweighting iterations $I_{\text{WPE}} = 10$, weight regularization parameter $\epsilon = 10^{-7}$ and sparsity-promoting parameter $p \in \{0.05,0.5,0.9\}$. The microphone-dependent prediction delays were computed with a prediction delay of $\tau = 2$ for the reference microphone, estimated time-differences-of-arrival using the generalised cross-correlation with phase transform (GCC-PHAT) \cite{Knapp1976} method and implemented using cross-band filtering\cite{Lohmann2023}.
\subsection{Simulation results}
For the WPE algorithm, we consider the following reference microphone selection methods:
\begin{itemize}
    \item $\ell_p$: reference microphone selection using \eqref{ref:lp_reference_selection_broadband} based on the $\ell_p$-norm of the dereverberated output with $I = I_{\text{WPE}}$ WPE iterations
    \item $\ell_p/\ell_2$: reference microphone selection using \eqref{ref:normalized_lp_norm_problem} based on the $\ell_p/\ell_2$-norm of the dereverberated output with $I = \{0,1,I_{\text{WPE}}\}$ WPE iterations
\end{itemize}
\begin{itemize}
    \item $\text{Max-}\hat{\text{ELR}}$: reference microphone selection by choosing the reverberant microphone signal with the largest estimated ELR using the method proposed in \cite{Xiong2019}
    \item $\text{Max-}\text{Power}$: reference microphone selection by choosing the reverberant microphone signal with the largest average signal power \cite{LawinOre2012}\cite{Araki2018}
\end{itemize}
It should be noted that the (normalized) $\ell_p$-norm-based reference microphone selection methods with $I > 0$ require running WPE in each reference microphone, whereas the reference microphone selection methods based on the normalized $\ell_p$-norm with $I = 0$ and based on the estimated ELR and signal power do not require any WPE iterations.


In order to compute the performance improvement of the considered reference microphone selection methods, the dereverberation performance, i.e. the quality of the dereverberated output, using the selected reference microphone is evaluated against the average dereverberation performance of all possible reference microphones, i.e.
\begin{equation}
\Delta\text{PESQ} = \text{PESQ}\big(\hat{d}_{\hat{r}}(t), s_{\hat{r}}(t)\big) - \text{PESQ}_{\text{avg}},
\label{eq:performance_improvement}
\end{equation}
where $\Delta\text{PESQ}$ denotes the perceptual evaluation of speech quality (PESQ) improvement with $\text{PESQ}\big(\hat{d}_{\hat{r}}(t), s_{\hat{r}}(t)\big)$ and $\text{PESQ}_{\text{avg}} = \frac{1}{M}\sum_{r = 1}^M\text{PESQ}\big(\hat{d}_{r}(t), s_{r}(t)\big)$ denoting the PESQ of the (time-domain) dereverberated output $\hat{d}_{\hat{r}}(t)$ with time index $t$ in selected reference microphone $\hat{r}$ and the average PESQ using all possible reference microphones, respectively. The target signal is the (time-domain) direct speech received at the reference microphone position $s_{r}(t)$. The improvement in the frequency-weighted segmental signal-to-noise ratio ($\Delta\text{FWSSNR}$) is defined similarly as in \eqref{eq:performance_improvement}.
The above measures \cite{Kinoshita2016} are averaged across the 12 considered positions of the speech source.
\par
For the considered reference microphone selection methods\footnote{Audio examples available on \\uol.de/f/6/dept/mediphysik/ag/sigproc/audio/dereverb/wpe-refmic-selection.html}, Fig. \ref{fig:performance} depicts the average performance improvement over all considered source positions in terms of $\Delta$FWSSNR and $\Delta$PESQ using a sparsity-promoting parameter $p = 0.05$, $p = 0.5$ and $p=0.9$. For all considered values of the sparsity-promoting parameter $p$, the performance of the proposed method using $I = I_{\text{WPE}}$ WPE iterations is larger than the performance using a selection based on the estimated ELR or signal power. Furthermore, a similar performance can be achieved using the proposed method with only $I = 1$ WPE iteration. Even when $I = 0$, the performance using the proposed method is similar to the performance of the considered estimated ELR and signal power-based reference microphone selection methods. Finally, it can be seen that the normalization of the dereverberated output is required for WPE reference microphone selection as the performance using the $\ell_p$-norm-based reference microphone selection method is significantly lower than the performance using the proposed method.
\begin{figure}[H]
\hspace{1.2cm}\includegraphics[width=0.75\columnwidth]{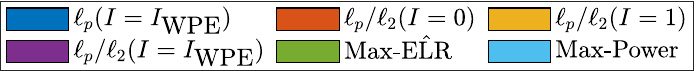}
\end{figure}%
\vspace*{-0.55cm}
\begin{figure}[H]%
\centering
\begin{subfigure}[H]{0.85\columnwidth}
  \includegraphics[width=\columnwidth]{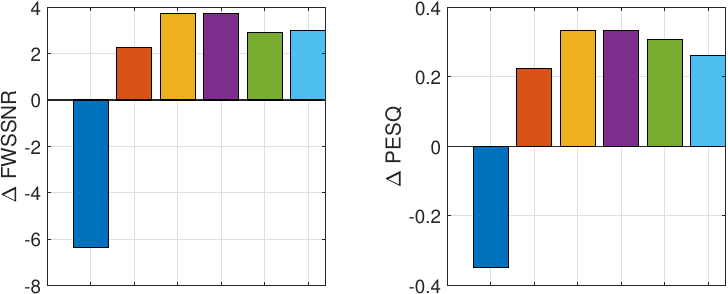}
  \vspace{-0.3cm}
  \caption{$p = 0.05$}%
  \label{subfig: sdn}%
\end{subfigure}\hfill%
\begin{subfigure}[H]{0.85\columnwidth}
  \includegraphics[width=\columnwidth]{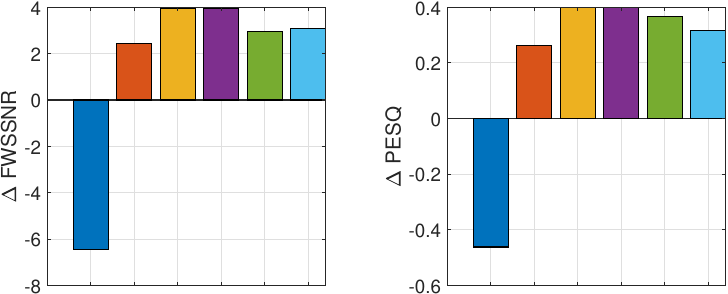}
  \vspace{-0.3cm}
  \caption{$p = 0.5$}%
  \label{subfig: sdb}%
\end{subfigure}
\begin{subfigure}[H]{0.85\columnwidth}
  \includegraphics[width=\columnwidth]{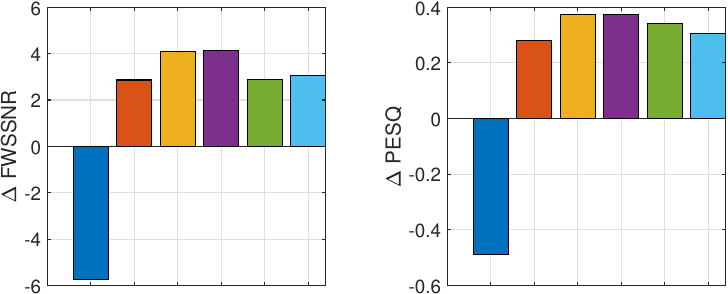}
  \vspace{-0.3cm}
  \caption{$p = 0.9$}%
  \label{subfig: sdb}%
\end{subfigure}
\vspace{-0.2cm}
\caption{Average performance improvement in terms of FWSSNR improvement and PESQ improvement for all considered reference microphone selection methods using a sparsity-promoting parameter (a) $p = 0.05$, (b) $p = 0.5$ and (c) $p=0.9$}
\label{fig:performance}
\vspace{-0.4cm} 
\end{figure}
\section{Conclusion}
\label{sec:majhead}
In this paper we have presented a reference microphone selection method for the WPE algorithm. Based on the WPE cost function, we first defined the reference microphone selection as an $\ell_p$-norm minimization problem. However, when considering spatially distributed microphones, the differences in signal power between the microphones may be large and the $\ell_p$-norm-based reference microphone selection problem may not yield the dereverberated output with the highest signal quality. Hence, we proposed to normalize for the power in the dereverberated output, leading to reference microphone selection based on the normalized $\ell_p$-norm. The experimental results showed that the performance of the proposed method is larger than the performance using a selection based on the estimated ELR or signal power. Furthermore, similar performance can be achieved for the proposed method using only a small number of WPE iterations. Investigating the performance of the considered reference microphone selection methods for acoustic scenarios with additive noise, a moving source or different microphone sensitivities are directions for future research.

\newpage

\bibliographystyle{IEEEbib}
\bibliography{refs}



\end{document}